\documentclass{an}
\usepackage{graphicx}
\usepackage{times}
\Volume{01}              
\Year{2005}              
\Month{01}               
\Pagespan{1}{6}      	 

\begin{document}
\lhead[\thepage]{A.N. Karaali: Parametrization of the Galactic by two exponentials}
\rhead[Astron. Nachr./AN~{\bf XXX} (200X) X]{\thepage}
\headnote{Astron. Nachr./AN {\bf 32X} (200X) X, XXX--XXX}

\title{Parametrization of the Galactic Structure by two exponentials}

\author{S. Karaali   
}
\institute{Istanbul University Science Faculty, 
           Department of Astronomy and Space Sciences, 
           34119, University-Istanbul, Turkey}
\date{} 

\abstract{We parametrized the total structure of the Galaxy in cylindrical 
coordinates by radial and vertical exponentials up to $z\sim10$ kpc, 
covering thin disc, thick disc, and the inner spheroid. However, we let the 
scaleheight and scalelength to be a continuous function of distance from the 
Galactic plane. The standard deviations for the differences between the space 
densities estimated by means of the newly defined scaleheight and scalelength 
and the observed space densities for three absolute magnitude intervals, 
$5<M(g)\leq6$, $6<M(g)\leq7$, and $7<M(g)\leq8$, for the fields 
SA 114, ELAIS, and \#0952+5245 are rather small. The uncertainties for the 
scaleheight are also small, indicating that this parameter is very sensitive 
to the distance from the Galactic plane, whereas those for the scalelength 
are larger.    
}
\correspondence{karsa@istanbul.edu.tr}

\maketitle

\section{Introduction}
For some years, a disagreement exists  among the researchers about  the 
formation  history of our Galaxy. Yet there has been a large improvement 
about this topic since the pioneering work of Eggen, Lynden-Bell \& Sandage (1962) 
who argued that the Galaxy collapsed in a free-fall time 
($\sim 2\times10^{8}$ yr). Now, we know that the Galaxy collapsed over many Gyr 
(e.g. Yoshii \& Saio 1979; Norris, Bessell, \& Pickles 1985; Norris 1986; Sandage 
\& Fouts 1987; Carney, Latham, \& Laird 1990; Norris \& Ryan 1991; Beers \& 
Sommer-Larsen 1995) and at least some of its components are formed from the merger 
or accretion of numerous fragments, such as dwarf-type galaxies (cf. Searle \& Zinn 
1978; Freeman \& Bland-Hawthorn 2002, and references therein). Also, the number of 
population components of the Galaxy increased  by one,  complicating 
interpretations of any data set. The new component (the thick disc) was introduced 
by Gilmore \& Reid (1983) in order to explain the observation that star counts 
towards the South Galactic Pole were not in agreement with a single-disc (thin-disc) 
component, but rather could be much better represented by two such components. This 
was the simplest combination of free parameters giving a satisfactory fit, and 
simplicity is generally key in astrophysical fits. 

Different parametrization followed the work of Gilmore \& Reid (1983). For example, 
Kuijken \& Gilmore (1989) showed that the vertical structure of our Galaxy could 
be best explained  by a multitude of quasi-isothermal components, i.e. a large 
number of sech$^{2}$ isothermal discs, together making up a more sharply peaked 
sech or exponential distribution. Also, we quote the work of de Grijs, Peletier \& van 
der Kruit (1997) who made clear that in order to build up a sech distribution, one 
needs multiple components.  Although different parametrization were tried by many 
reserachers, only the one of Gilmore \& Wyse (1985) which is based on star counts 
estimation for thin disc, thick disc and spheroid (halo) became as a common model 
for our Galaxy, and used widespread with improving the parameters, however. In other words, 
 the canonical density laws are as follows: 1) a parametrization for thin and thick discs 
in cylindrical coordinates by radial and vertical exponentials and 2) a parametrization 
for halo by the de Vaucouleurs (1948) spheroid. The thin disc dominates the small $z$ 
distances from the galactic plane with a scaleheight ranging from 200 to 475 pc 
(Robin \& Cr\'{e}z\'{e} 1986), whereas the thick disc extends to larger $z$ distances 
with larger scaleheight. In some studies, the range of values for the parameters is 
large, especially for the thick disc. For example, Chen et al. (2001) and Siegel et 
al. (2002) give 6.5-13 and 6-10 per cent, respectively, for the relative local density 
for the thick disc. In the paper of Karaali, Bilir \& Hamzao\u glu (2004), we 
discussed the large range of these parameters and claimed that Galactic model parameters 
are absolute magnitude dependent. We showed that the range of the model parameters 
estimated for a unique absolute magnitude interval is considerably smaller. 

In  the present study  different procedure is followed. We show that 
logarithmic space densities can be parametrized by two exponentials for each absolute 
magnitude interval. However, we let the scaleheight and the scalelength to be a 
continuous function of the distance from the galactic plane. Also, we show that the 
efficiency of the thick disc is absolute magnitude dependent.

In Sections 2 and 3, the canonical density law forms and the new procedure for 
density evaluation is discussed. The calibration of the scaleheight and scalelength 
for three absolute magnitude intervals for three fields is given in Section 4. 
 In  Section 5 the densities  are compared  at different distances from 
the galactic plane evaluated by the new calibration for three fields. Finally 
Section 6 provides a discussion.
    	       
\section{The canonical density law forms}
Disc structures are usually parameterized in cylindrical coordinates by radial 
and vertical exponentials,

\begin{eqnarray}
\tiny
D_{i}(x,z)=n_{i} exp(-z/H_{i}) exp(-(x-R_{o})/h_{i})
\end{eqnarray}
where $z$ is the distance from galactic plane, $x$ is the planar distance 
from the Galactic center, $R_{0}$ is the solar distance to the Galactic 
center (8.6 kpc), $H_{i}$ and $h_{i}$ are the scaleheight and scalelength 
respectively, and $n_{i}$ is the normalized local density. The suffix $i$ 
takes the values 1 and 2, as long as the thin and thick discs are considered. 
A similar form uses the sech$^{2}$ (or sech) function to parametrize 
the vertical distribution for the thin disc,

\begin{eqnarray}
D_{1}(x,z)=n_{1}sech^{2}(-z/H^{'}_{1})exp(-(x-R_{o})/h_{1}).
\end{eqnarray}
Because the sech function is the sum of two exponentials, $H^{'}_{1}$ is not really 
a scaleheight, but has to be compared to $H_{1}$ by dividing it with 2: 
$H_{1}=H_{1}^{'}/2$ (van der Kruit \& Searle 1981a, b, 1982a, b; van der Kruit 1988).
However, in order to build up a sech distribution, one needs multiple components 
(Kuijken \& Gilmore 1989, de Grijs et al. 1997). 

The density law for the spheroid (halo) component is parameterized in 
different forms. The most common is the de Vaucouleurs (1948) spheroid used 
to describe the surface brightness profile of elliptical galaxies. This law has 
been deprojected into three dimensions by Young (1976) as 

\begin{eqnarray}
D_{s}(R)=n_{s}~exp[-7.669(R/R_{e})^{1/4}]/(R/R_{e})^{7/8},
\end{eqnarray}
where $R$ is the (uncorrected) Galactocentric distance in spherical 
coordinates, $R_{e}$ is the effective radius and $n_{s}$ is the normalized 
local density. $R$ has to be corrected for the axial ratio $\kappa = c/a$, 

\begin{eqnarray}
R = [x^{2}+(z/\kappa)^2]^{1/2},
\end{eqnarray}
where,
\begin{eqnarray}
x = [R_{o}^{2}+(z/\tan b)^2-2R_{o}(z/\tan b)\cos l]^{1/2},
\end{eqnarray}
$b$ and $l$ being the Galactic latitude and longitude, respectively, for the 
field under investigation. 

An alternative formulation is the power law,
\begin{eqnarray}
D_{s}(R)=n_{s}/(a_{o}^{n}+R^{n})
\end{eqnarray}
where $a_{o}$ is the core radius.

If one restricts the work  to  vertical direction, then the third factor in 
eqs. (1) and (2)  can be neglected.  

\section{Unique equation for space densities}
We mentioned in Section 1 that the introduction of the thick disc component 
into the literature was due to better representation of the observed star counts 
in the South Galactic Pole. Principally, what we are doing is  matching  
the observed data with the appropriate estimated data. Hence, we can modify and 
limit the density law forms cited in Section 2 for this purpose. We  can  
argue that the total structure of our Galaxy can be parametrized in cylindrical coordinates 
by radial and vertical exponentials in the form of eq. (1) up to many kiloparsecs, 
covering thin and thick discs and probably the inner spheroid. However, we let 
the scaleheight and the scalelength to be a continuous function of distance from 
the galactic plane. Our  claim  is based on the following:
\begin{enumerate}
\item The density law forms for thin and thick discs are similar. There may be 
a continuous transition for the scaleheight from short distances to large 
distances in our Galaxy. In the canonical way, we estimate two scaleheights for 
two discs which are the mean values for relatively short and large distances. 
The same case is valid for the scalelength.
\item Although there is a tendency in the recent works (cf. Feltzing, Bensby, 
\& Lundstr\"om, 2003) that the two discs are discrete components, their 
kinematical data and $[Fe/H]$ metallicities overlap.  
\item There is almost a consensus for the double structure of the spheroid, 
inner and outermost spheroids. According to Norris (1996), there are a number 
of indications that a significant fraction of material with $[Fe/H]<-1$ has 
disclike signature. Hence, the density law form for inner spheroid may be 
similar to the density law form of discs.
\end{enumerate}

The most important point at this approach is the calibration of the scaleheight 
$H$ and scalelength $h$ with the distance $z$ from the galactic plane. We followed 
the following procedure for this purpose. First we write eq. (1) in the 
logarithmic form:

\begin{eqnarray}
D^{*}(x,z)=n^{*}-(\frac{z}{H} + \frac{x-R_{o}}{h})\log ({e})
\end{eqnarray}
where $D^{*}(x,z)=\log D(x,z) + 10$ and $n^{*}=n+10$. If $x$ of (5) is substituted 
in eq. (7) and re-written, it takes the form

\begin{eqnarray}
n^{*}-D^{*}(z,l,b)=\{(z/H)+(1/h)[R_{0}^{2}+(z/\tan b)^{2}\nonumber\\
-2R_{o}(z/\tan b)\cos l]^{1/2}-R_{0}]\}\log ({e})~~~~
\end{eqnarray}
Now, we can use the Hipparcos' local space density (Jahreiss \& Wielen, 1997) for 
a specific absolute magnitude interval $M_{1}-M_{2}$ and a sequence of observed 
$D^{*}(z, l, b)$ space densities and estimate the most likely $H$ and $h$ values, 
for each element of this sequence.   

\begin{table}
\center
\caption{Galactic coordinates (epoch 2000) and the sizes for three fields investigated.}   
\begin{tabular}{lccc}
\hline
Field      &	l 		&    b	         & Size (deg$^{2}$)\\
\hline
SA 114     & 	$68^{o}.50$	& $-48^{o}.38$   & 	 4.239\\
ELAIS 	   &	$84^{o}.27$	& $+44^{o}.90$   &	 6.571\\
\#0952+5245& 	$83^{o}.38$	& $+48^{o}.55$   &	20\\
\hline
\end{tabular}   
\end{table}

\begin{figure*}
\center
\resizebox{11.5cm}{13.66cm}{\includegraphics*{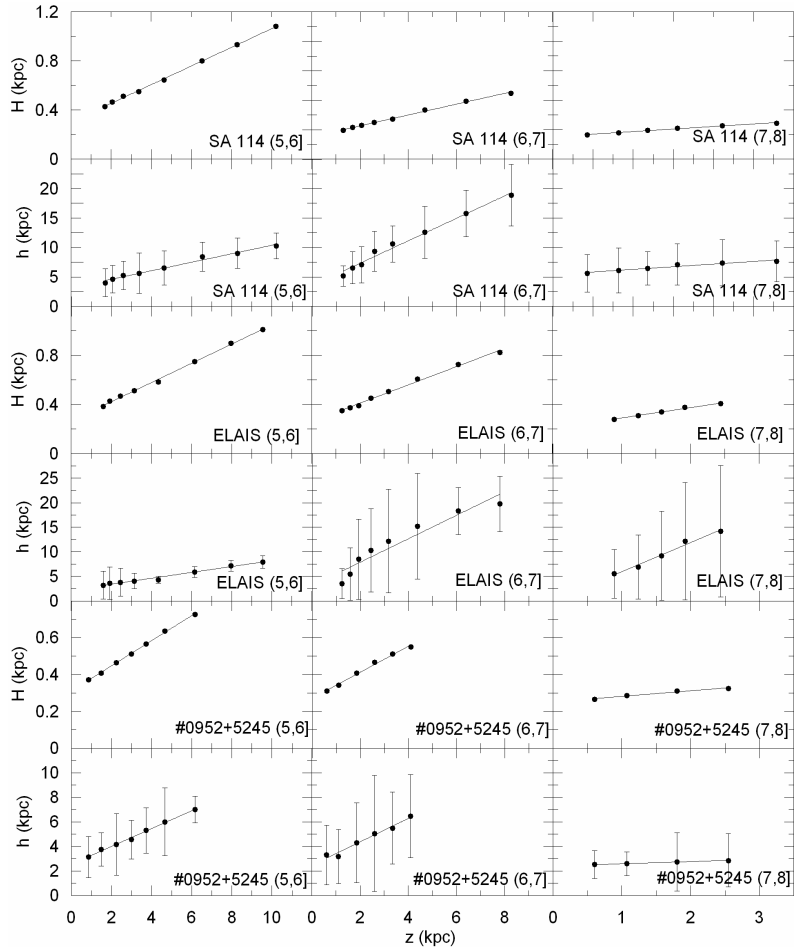}} 
\caption {Calibration of the scaleheight and scalelength with distance from the 
galactic plane for the absolute magnitude interval (5,6], (6,7], and (7,8], for 
the fields SA 114, ELAIS, and \#0952+5245. The errors bars for the
scaleheight could not be demonstrated, since they are small (see Table 2).}
\end {figure*}

\begin{table*}
\center
{\scriptsize
\caption{The most likely scaleheight (H) and scalelength (h) values for a given 
distance from the galactic plane ($z^{*}$) and the corresponding logarithmic 
space density $D^{*}$, for the absolute magnitude intervals (5,6], (6,7], and 
(7,8], for the  fields SA 114, ELAIS, and \#0952+5245. The local space density 
of Hipparcos are $n^{*}=7.47$, for absolute magnitude intervals (5,6], (6,7] and 
$n^{*}=7.48$, for (7,8].}   
\begin{tabular}{lrrrrrrrrrrrr}
\hline
$M(g) \rightarrow$ & \multicolumn{4} {c} {(5,6]} & \multicolumn{4} {c} {(6,7]} & \multicolumn{4} {c} {(7,8]}\\
\hline
Field & $z^{*}$ (kpc) & $D^{*}$ & H (kpc) & h (kpc)& $z^{*}$ (kpc)& $D^{*}$ & H (kpc)& h (kpc) & $z^{*}$ (kpc)& $D^{*}$ & H (kpc) & h (kpc)\\
\hline
SA 114&1.69 &5.68 & 0.43$\pm$0.007& 4.00$\pm$2.38& 1.29 & 6.00 & 0.39$\pm$0.012& 5.15$\pm$1.76& 0.93 & 6.23 & 0.33$\pm$0.009&5.60$\pm$3.15\\
&2.06 &5.46 & 0.46~~0.013& 4.61~~2.34& 1.68 & 5.73 & 0.43~~0.009& 6.55~~2.71& 1.32 & 5.83 & 0.36~~0.009&6.10~~3.82\\
&2.60 &5.17 & 0.51~~0.007& 5.24~~2.43& 2.06 & 5.47 & 0.46~~0.008& 7.05~~3.07& 1.68 & 5.57 & 0.39~~0.005&6.44~~2.83\\
&3.38 &4.69 & 0.55~~0.009& 5.60~~3.43& 2.59 & 5.17 & 0.50~~0.006& 9.35~~3.39& 2.05 & 5.31 & 0.42~~0.008&7.10~~3.49\\
&4.63 &4.21 & 0.65~~0.004& 6.51~~2.90& 3.35 & 4.73 & 0.54~~0.011&10.59~~3.09& 2.61 & 4.93 & 0.46~~0.004&7.39~~3.89\\
&6.53 &3.75 & 0.80~~0.009& 8.45~~2.48& 4.68 & 4.35 & 0.67~~0.006&12.57~~4.41& 3.29 & 4.47 & 0.49~~0.006&7.65~~3.44\\
&8.28 &3.39 & 0.93~~0.012& 9.01~~2.60& 6.40 & 3.86 & 0.79~~0.008&15.74~~3.95&      &      &        &              \\
&10.21&3.11 & 1.08~~0.012&10.26~~2.19& 8.27 & 3.36 & 0.90~~0.007&18.85~~5.22&      &      &        &              \\
            &     &      &      &      &      &      &      &       &      &      &      &                        \\
ELAIS &1.60 & 5.66 & 0.39$\pm$0.002 & 3.20$\pm$2.83 & 1.24 & 5.94 & 0.35$\pm$0.007 &  3.50$\pm$3.06 & 0.89 & 6.10 & 0.28$\pm$0.006 &5.50$\pm$4.94\\
&1.93 & 5.51 & 0.43~~0.005 & 3.60~~3.31 & 1.58 & 5.63 & 0.37~~0.010 &  5.46~~5.35 & 1.24 & 5.75 & 0.31~~0.007 &  6.90~~6.51\\
&2.46 & 5.17 & 0.47~~0.003 & 3.80~~2.85 & 1.94 & 5.31 & 0.39~~0.006 &  8.50~~8.20 & 1.58 & 5.46 & 0.34~~0.006 &  9.20~~9.08\\
&3.15 & 4.77 & 0.51~~0.008 & 4.04~~1.57 & 2.45 & 5.11 & 0.45~~0.006 & 10.30~~8.46 & 1.92 & 5.26 & 0.38~~0.003 & 12.15~~11.95\\
&4.35 & 4.17 & 0.58~~0.009 & 4.32~~0.73 & 3.18 & 4.73 & 0.51~~0.003 & 12.20~~10.55& 2.44 & 4.87 & 0.41~~0.003 & 14.20~~13.40\\
&6.15 & 3.80 & 0.75~~0.007 & 5.90~~1.07 & 4.38 & 4.32 & 0.61~~0.002 & 15.20~~10.76&      &      &      &           \\
&7.97 & 3.47 & 0.90~~0.008 & 7.17~~1.12 & 6.07 & 3.80 & 0.72~~0.013 & 18.30~~4.80 &      &      &      &           \\
&9.54 & 3.17 & 1.01~~0.009 & 7.93~~1.27 & 7.81 & 3.30 & 0.82~~0.014 & 19.76~~5.64 &      &      &      &           \\
      &     &      &      &      &      &      &      &       &      &      &      &       \\
\#0952+5245&0.86 & 6.47 & 0.37$\pm$0.004&3.14$\pm$1.68& 0.61 & 6.62 & 0.31$\pm$0.006& 3.30$\pm$2.42 & 0.60 & 6.50 & 0.27$\pm$0.005 &  2.52$\pm$1.15\\
&1.48 & 5.90 & 0.41~~0.006 & 3.75~~1.36 & 1.10 & 6.08 & 0.34~~0.004 &3.17~~2.20&1.08& 5.86 & 0.29~~0.005 &2.60~~0.97\\
&2.25 & 5.36 & 0.46~~0.006 & 4.16~~2.51 & 1.86 & 5.50 & 0.41~~0.002 &4.30~~3.25&1.80& 4.97 & 0.31~~0.004 &2.74~~2.36\\
&2.99 & 4.93 & 0.51~~0.004 & 4.55~~1.58 & 2.59 & 5.05 & 0.47~~0.002 &5.05~~4.73&2.55& 4.07 & 0.32~~0.005 &2.86~~2.17\\
&3.73 & 4.59 & 0.57~~0.004 & 5.30~~1.85 & 3.34 & 4.62 & 0.51~~0.005 &5.50~~2.93&      &      &        &       \\
&4.66 & 4.26 & 0.64~~0.005 & 6.00~~2.77 & 4.10 & 4.21 & 0.55~~0.005 &6.46~~3.38&      &      &        &       \\
&6.17 & 3.72 & 0.73~~0.006 & 7.01~~1.08 &      &      &             &          &      &      &        &       \\
\hline
\end{tabular}
}  
\end{table*}

\section {Absolute magnitude dependent scaleheight and scalelength as a function 
of distance from the galactic plane}

The eq. (8) applied to the data of three fields whose galactic coordinates and 
sizes are given in Table 1. Fields SA 114 and ELAIS are almost symmetric relative 
to the galactic plane, and the galactic coordinates of the centre of ELAIS and 
\#0952+5245 almost coincide with one another, however their sizes are quite 
different. 

For the field SA 114, we transfered the mean $z^{*}$ distances from the galactic 
plane and the corresponding logarithmic space densities $D^{*}$ for the absolute 
magnitude intervals $5<M(g^{'})\leq6$, $6<M(g^{'})\leq7$, and $7<M(g^{'})\leq8$ 
from the work of Karaali et al. (2004), whereas the $z^{*}$ and 
$D^{*}$ data for the same absolute magnitude intervals, for the fields ELAIS and 
\#0952+5245 were recently evaluated by Bilir, Karaali \& Gilmore (2005) and Karaali 
et al. (2005), respectively. The space densities for these intervals extend up to 
$\sim$10 kpc from the galactic plane and cover thin and thick discs, and 
spheroid. Here, $D^{*}=\log D+10$, $D=N/ \Delta V_{1,2}$; 
$\Delta V_{1,2}=(\pi/180)^{2}(\sq/3)(r_{2}^{3}-r_{1}^{3})$; $\sq$ denotes the size 
of the field; $r_{1}$ and $r_{2}$ denote the limiting distance of the volume 
$\Delta V_{1,2}$; and $N$ denotes the number of stars in this volume. The sizes 
of the fields SA114, ELAIS, and \#0952+5254 are 4.239, 6.751, and 20 deg$^{2}$ 
respectively (see also Table 1). The most likely scaleheights and scalelengths for 
these intervals are given in Table 2, and the calibrations of scaleheights and 
scalelengths in this table with the distance from the galactic plane are shown in 
Fig. 1. We used the procedure of Phleps et al. (2000) for the error estimation in 
Table 2 and Fig. 1, i.e. changing the values of the parameters until $\chi^{2}$ 
increases or decreases by 1. The errors for the scaleheight are rather small, 
indicating that this parameter is  a strong function of  the distance from 
the galactic plane. Whereas, the errors for the scalelenght are large which show 
that the scalelenght is not effective in the comparison of observed logarithmic 
space density and the estimated one in eq. (7). The correlation is higher for the 
calibration of scaleheight for all absolute magnitudes and for all fields. 
We adopted  the following linear equations in the calibrations and this 
resulted in the fluctuations appearing on the figures to be smoothed.  

\begin{eqnarray}
H = a_{1}z+a_{0}
\end{eqnarray}
\begin{eqnarray}
h = b_{1}z+b_{0}
\end{eqnarray}

The coefficients for the eqs. (9) and (10), evaluated by least-square method, are 
given in Table 3 together with their rather small uncertainties. All the 
scaleheights and scalelengths are increasing functions of the distance from 
the galactic plane. Hence, the scaleheight and scalelength increases from short $z$ 
distances to the large ones. The numerical values of the scaleheight for short $z$ 
distances are in the range of the scaleheight for thin disc claimed so far. For 
example, the scaleheight for the absolute magnitude interval $7<M(g)\leq8$ for the 
field \#0952+5245 lies within 0.27 and 0.32 kpc. However, the upper limits for the 
scaleheights for the same absolute magnitude interval, for the fields SA 114 and 
ELAIS are larger, i.e. 0.49 and 0.41 kpc, respectively. For larger $z$ distances, 
the numerical value of scaleheight is close to the scaleheight of thick disc 
appeared in the literature. For example, $H\sim0.5$ kpc at the distance from the 
galactic plane $z\sim 3.5$ kpc, for three fields. The scaleheight extends up to 
$H\sim 1$ kpc at $z\sim10$ kpc. The range of the scalelength is rather large, and 
it  differs  from field to field. The least range as well as the least numerical 
values belong to the absolute magnitude interval $7<M(g)\leq8$ for \#0952+5245, 
$2.5<h<2.9$ kpc, and the largest ones belong to $6<M(g)\leq7$ for ELAIS field, 
$3.5<h<19.8$ kpc. Such large scalelengths have not been claimed in the literature 
up to now.

\begin{table}
\center
\caption{Coefficients for eqs. (9) and (10) for three absolute magnitude intervals 
for the three fields.}
{\scriptsize
\begin{tabular}{ccccc}
\hline
M(g) &    $a_{1}$ &    $a_{0}$ &    $b_{1}$ &    $b_{0}$ \\
\hline
SA 114&            &            &            &            \\
(5,6] & 0.0760$\pm$0.0007 & 0.3020$\pm$0.0041& 0.7210$\pm$0.0385 & 3.1607$\pm$0.2205\\
(6,7] & 0.0734~~~0.0020   & 0.3068~~~0.0090  & 1.9057~~~0.0953   & 3.5076~~~0.4235  \\
(7,8] & 0.0688~~~0.0040   & 0.2704~~~0.0085  & 0.8906~~~0.1166   & 4.9499~~~0.2487  \\
ELAIS &            &            &            &            \\
(5,6] & 0.0784$\pm$0.0018& 0.2654$\pm$0.0094 & 0.5976$\pm$0.0310 & 2.2205$\pm$0.1677\\
(6,7] & 0.0744~~~0.0026  & 0.2621~~~0.0110   & 2.3905~~~0.2977   & 3.0923~~~1.2495  \\
(7,8] & 0.0837~~~0.0054  & 0.2073~~~0.0092   & 5.9479~~~0.4658   &-0.0100~~~0.7923  \\
\#0952+5245&       &            &            &            \\
(5,6] & 0.0683$\pm$0.0013& 0.3108$\pm$ 0.0047& 0.7261$\pm$0.0686 & 2.5468$\pm$0.2468\\
(6,7] & 0.0707~~~0.0038  & 0.2713~~~0.0099   & 0.9526~~~0.0803   & 2.4696~~~0.2066  \\
(7,8] & 0.0303~~~0.0041  & 0.2515~~~0.0068   & 0.1769~~~0.0050   & 2.4134~~~0.0083  \\
\hline
\end{tabular}
}  
\end{table}

\begin{table}
\center
\caption{Standard deviations for the differences between the observed and 
evaluated logarithmic space densities  for three absolute magnitude intervals, 
for three fields.} 
\begin{tabular}{cccc}
\hline
Field &     SA 114 &      ELAIS  &\#0952+5245 \\
M(g)  &        $s$ &        $s$  &        $s$ \\
\hline
(5,6] & $\pm$ 0.03 &  $\pm$ 0.06 & $\pm$ 0.02 \\
(6,7] & $\pm$ 0.05 &  $\pm$ 0.07 & $\pm$ 0.04 \\
(7,8] & $\pm$ 0.04 &  $\pm$ 0.03 & $\pm$ 0.03 \\
\hline
\end{tabular}  
\end{table}

\begin{table*}
\center
\caption{Comparison of the logarithmic space densities ($D^{*}$) evaluated for 
a sequence of distance from the galactic plane ($z$), for three absolute 
magnitude intervals, for three fields.} 
\begin{tabular}{ccccccccccc}
\hline
$M(g) \rightarrow$ & \multicolumn{3} {c} {(5,6]} & \multicolumn{3} {c} {(6,7]} & \multicolumn{4} {c} {(7,8]}\\ 
$z$ (kpc) & SA 114 & ELAIS &\#0952+5245 & SA 114 & ELAIS &\#0952+5245 & $z$ (kpc) & SA 114 &ELAIS &\#0952+5245\\
\hline
0 & 7.47 & 7.47 & 7.47 & 7.47 & 7.47 & 7.47 & 0 & 7.48 & 7.48 & 7.48 \\
1 & 6.28 & 6.21 & 6.33 & 6.30 & 6.18 & 6.21 & 1 & 6.17 & 5.99 & 5.95 \\
2 & 5.48 & 5.41 & 5.53 & 5.51 & 5.35 & 5.37 & 2 & 5.30 & 5.16 & 4.70 \\
4 & 4.47 & 4.42 & 4.47 & 4.50 & 4.35 & 4.31 & 3 & 4.67 & 4.63 & 3.66 \\
6 & 3.85 & 3.82 & 3.79 & 3.89 & 3.76 &      &   &      &      &      \\
8 & 3.44 & 3.42 &      & 3.48 & 3.37 &      &   &      &      &      \\
10 &3.14 & 3.12 &      &      &      &      &   &      &      &      \\
\hline
\end{tabular}  
\end{table*}

\section{Testing the new calibrations}
We replaced the calibrations of scaleheights and scalelengths into eq. (8), and 
we evaluated the logarithmic space densities, $D^{*}$, for the $z^{*}$ distances 
for which observed space densities are available in Table 2 for three fields. 
Then, we compared them with the observed space densities. The 
standard deviations for differences between the observed and evaluated space 
densities are small (Table 4), confirming the new calibration, i.e. the total 
structure of the Galaxy can be parametrized in cylindrical coordinates by radial 
and vertical exponentials up to many kiloparsecs ($\sim$ 10 kpc), covering the 
thin disc, the thick disc, and the halo. However, contrary to the procedures in 
situ, the scaleheight and scalelength are not constants, but they are continuous 
functions of distance from the galactic plane. 

We have a parametrization for each absolute magnitude interval, for each field. 
Now, a question: Do the parametrizations for a specific absolute magnitude interval 
for three fields produce the same space density for a given distance from the 
galactic plane? The answer  to  this question can be  obtained  from 
Table 5 where logarithmic space densities are given for three absolute magnitude 
intervals for three fields. For the absolute magnitude interval $5<M(g)\leq6$, the field 
\#0952+5245 is investigated only up to $z\sim6$ kpc, whereas the space densities 
for the fields SA 114, and ELAIS extend up to $z\sim10$ kpc. The logarithmic space 
densities for three fields are rather close to each other for a given $z$. For the 
absolute magnitude interval $6<M(g)\leq7$, the agreement of the logarithmic space 
densities is also good for the fields \#0952+5245 and ELAIS, however it is a bit 
less for SA 114. The logarithmic space densities for $7<M(g)\leq8$ extends up to 
only $z\sim3$ kpc, and it is interesting, that the agreement is better between the 
data of SA 114 and ELAIS fields.        

\section{Discussion}

The parametrization of the Galactic components has a long history. Bahcall \& 
Soneira (1980) parametrized the disc in cylindrical coordinates by radial and 
vertical exponentials, and the halo by a de Vaucouleurs (1948) spheroid in 
their two component Galactic model. Whereas, Gilmore \& Reid (1983) introduced 
a new component, thick disc, in order to explain their star count observations. 
Later, it was noticed that the third component was a rediscovery of Intermediate 
Population II which was named in the Vatican conference (O'Connell, 1958). 

Different parametrizations followed the work of Gilmore \& Reid (1983). For 
example, Kuijken \& Gilmore (1989) showed that the vertical structure of our 
Galaxy could be best explained by a multitude of quasi-isothermal components, 
i.e. a large number of sech$^{2}$ isothermal discs, together making up a more 
sharply peaked sech or exponential distribution. Also, we quote the work of de 
Grijs et al. (1997) who made clear that in order to build up a sech 
distribution, one needs multiple components.  Although various parametrization 
were tried by many reserachers, only the one of Gilmore \& Wyse (1985) became as 
a common model for our Galaxy. In other words,  the canonical density laws 
are as follows: 1) a parametrization for the thin disc, 2) a parametrization for 
the thick disc both in cylindrical coordinates by radial and vertical exponentials, 
and 3) a parametrization for the halo by the de Vaucouleurs (1948) spheroid. 

In some studies, the range of values for the parameters is large, especially for 
the thick disc. For example, Chen et al. (2001) and Siegel et al. (2002) give 6.5-13 
and 6-10 per cent, respectively, for the relative local density for the thick disc. 
In the paper of Karaali et al. (2004), we discussed the large range of these 
parameters and claimed that Galactic model parameters are absolute magnitude 
dependent. We showed that the range of the model parameters estimated for a unique 
absolute magnitude interval is considerably smaller.

In this paper, we used the results obtained in our previous paper, thus we estimated 
model parameters as a function of absolute magnitude. Additionally, we showed that 
the derived logarithmic space densities could be parametrized by two exponentials 
only, for each absolute magnitude interval, without regarding the population type 
of stars. However, we let the scaleheight and scalelength to be a continuous 
function of distance from the galactic plane. This is the main difference between 
the works of Kuijken \& Gilmore (1989), de Grijs et al. (1997) and our work. The 
starting point is that the structure of two discs are parametrized in the same way, 
i.e. by two exponentials, and that there are a number of indications that a 
significant fraction of material with $[Fe/H]<-1$ (halo material) has disclike 
signature (Norris 1996). Here we showed that the scaleheight and the scalelength 
could be calibrated with $z$ distance from the galactic plane by linear functions, 
and the space densities evaluated for the absolute magnitude intervals $5<M(g)\leq6$, 
$6<M(g)\leq7$, and $7<M(g)\leq8$ for three fields could be explained by two 
exponentials with these scaleheights and scalelengths. The (small) standard 
deviations corresponding to the differences between the observed logarithmic space 
densities and the evaluated ones by the calibrations (Table 4) strongly confirm our 
suggestion. The scaleheights corresponding to small $z$ distances are typical thin 
disc scaleheights, whereas those at larger $z$ distances are at the level of thick 
disc scaleheights (Table 2). For example, the scaleheight for stars with $7<M(g)\leq8$ 
at $z=0.6$ kpc for the field \#0952+5245 is $H=0.27$ kpc, and the one for stars with 
$5<M(g)\leq6$ at $z=4.66$ kpc is $H=0.64$ kpc. In Table 5, we compare the logarithmic space 
densities ($D^{*}$) evaluated by the new calibration for a sequence of distance 
from the Galactic plane ($z$), for three absolute magnitude intervals, for three 
fields. For the absolute magnitude interval $5<M(g)\leq6$, the $D^{*}$-values for 
a given $z$ are rather close to each other for three fields. The same argument 
holds for the interval $6<M(g)\leq7$, however the agreement between the data of 
ELAIS field and \#0952+5245 is better, and finally for $7<M(g)\leq8$ the agreement 
between the data of SA 114 and ELAIS favors. These small differences are real and 
they reflect the differences between the corresponding coefficients in Table 3. 
Actually, the numerical values for $a_{1}$ and $a_{0}$ for the absolute magnitude 
interval $5<M(g)\leq6$ for three fields are rather close to each other, whereas 
for the interval $6<M(g)\leq7$, $a_{0}$ is 0.26 and 0.27 for the fields ELAIS and 
\#0952+5245 respectively, but it is 0.31 for the field SA 114. From the other 
hand, for the interval $7<M(g)\leq8$, $a_{1}$ is 0.07 and 0.08 for the fields 
SA 114 and ELAIS, but it is 0.03 for the field \#0952+5245. We evaluated the 
number of stars per deg$^{2}$ for the absolute magnitude intervals $6<M(g)\leq7$ 
and $7<M(g)\leq8$ for three fields to  find out  its effect on the agreement between 
the space densities for a specific $z$ distance in question.  We found  that 
there is no any correlation between the number of stars per deg$^{2}$ and the 
space densities for any of the absolute magnitude interval claimed above, for 
three fields. The claimed small differences is not an unexpected case. It is 
worthwhile to emphasize that the Galactic model parameters determined, by means 
of the data in different directions of the Galaxy, in situ do not overlap either. 
Table 1 of Karaali et al. (2004) which gives the Galactic model parameters for 
recent works confirms our claim. We can give an additional example. Although the 
Galactic coordinates of the ELAIS field and \#0952+5245 are almost the same, the 
corresponding model parameters determined in situ for these fields are not the same 
(Table 6). Probably, the difference originates from their sizes (see Table 1) 
and/or from different absolute magnitude intervals by which the model parameters 
were estimated, i.e. $5<M(g)\leq10$ and $4<M(g)\leq9$ for the ELAIS field and for 
the field \#0952+5245, respectively.

\begin{table}
\center
\caption{Galactic model parameters estimated by the data of stars with $5<M(g)\leq10$ 
and $4<M(g)\leq9$, respectively, in fields ELAIS and \#0952+5245, by the procedure 
in situ. Symbols: $n^{*}$ is the logarithmic local space density, H is the scaleheight, 
$n/n_{1}$ is the local space density relative to the thin disc and $\kappa$ is the 
axial ratio for the halo.}
{\scriptsize
\begin{tabular}{ccccc}
\hline
Field &Parameter & Thin disc & Thick disc & Halo\\
\hline
ELAIS & $n^{*}$   & $7.51^{+0.03}_{-0.02}$ & $6.26^{+0.07}_{-0.07}$  & $4.48^{+0.37}_{-0.53}$ \\
      & $H$ (pc)  & $274^{+8}_{-6}$ & $807^{+78}_{-61}$ & $-$ \\
      & $n/n_{1}$ (per cent) & $-$ & $5.62$ & $0.09$ \\
      & $\kappa$  & $-$ & $-$& $0.65^{+0.35}_{-0.28}$\\
      &           &     &    &\\   
\#0952+5245&  $n^{*}$   & $7.43^{+0.04}_{-0.05}$ & $6.14^{+0.09}_{-0.11}$  & $4.16^{+0.22}_{-0.41}$ \\
  & $H$ (pc)  & $276^{+12}_{-12}$ & $895^{+114}_{-96}$ & $-$ \\
  & $n/n_{1}($per cent$)$ & $-$ & $5.13$ & $0.05$ \\
  & $\kappa$  & $-$ & $-$& $0.59^{+0.31}_{-0.21}$\\
\hline
\end{tabular}  
}
\end{table}

It is worthwhile to note that small-number statistics could affect the robustness 
of the results presented. One can notice immediately that the uncertainty for 
such statistics is large. For example, the standard deviation or its equivalent 
value, the probable error, is large due to the small number in the denominator of 
the expression defining it.

We can generalize this topic as such: In many areas of astronomy and astrophysics 
it occasionally happens that only small number of events of interest are detected 
during an observation. Examples range from the number of supernovae seen in a given 
period of time from a cluster of galaxies to the number of gamma rays detected 
during a source observation. If the goal is to determine quantities such as the 
event rate or the ratio of different event types, then the best approach would be to 
repeat the measurement with a longer integration time or a larger collection factor 
in order to obtain enough events for an accurate measurement. In some cases, for 
one reason or another, this is not possible or practical, and one is forced to make 
the best use of the data in hand. Results are then typically quoted as upper limits 
at a specified confidence level or as a measured value with error bars containing a 
specified confidence interval.

Here we pose a  second  question: Does this work refuse the multistructure of 
the Galaxy? The answer could be ``yes'' if this question was asked about fifteen 
years ago. Today, there is a consensus about the existence of the thick disc 
not only in our Galaxy but also in external galaxies. However, there are some 
points that remains to be explained. We direct the reader to the work of 
Feltzing et al. (2003) and the references therein. According to these authors, 
the ages of thin disc ($6.1\pm3.8$ Gyr) and thick disc ($12.1\pm2.0$ Gyr) do 
not overlap, indicating two discrete components for our Galaxy. However, the 
$[Fe/H]$ metal abundances for two disc overlap in the range $-0.8<[Fe/H]<-0.3$ 
(Nissen, 2003). Although the trend of $\alpha$-elements is different for two 
discs for $[Fe/H]<0$, giving a chance to separate thin and thick disc stars, 
this is not the case for $[Fe/H]\geq0$. According to Feltzing et al. (2003) 
and Bensby et al. (2004a, b), metal rich stars  abundant  in $\alpha$-elements 
can be explained as the long formation time of thick disc. There is overlapping 
also for kinematical data, as claimed by many authors. The ``transition 
objects'' claimed recently by Bensby et al. (2004b) is a good confirmation of 
this argument.

The points claimed in the previous paragraph indicate the relation between thin 
and thick discs. Different data of these components overlap. They are not isolated 
parts of our Galaxy, but complementary ones. Hence, we may adopt the same argument 
for the distribution of their space density, as we carried out in this work.

\begin{acknowledgements}
I acknowledge the referee Dr. Richard de Grijs, whose thoughtful and constructive 
comments greatly improved this work. Also, I thank Dr. Esat Hamzao\u glu for 
checking the English of the text and Dr. Sel\c{c}uk Bilir for drawing the figures 
and tables. 
\end{acknowledgements}


\begin{thebibliography}{}
\bibitem{}Bahcall, J.N., Soneira, R.M.: 1980, ApJS, 44, 73
\bibitem{}Beers, T.C., Sommer-Larsen, J.: 1995, ApJS, 96, 175
\bibitem{}Bensby, T., Feltzing, S., Lundstr\"om, I., Ilyin, I.: 2004a, A\&A, 410, 527 
\bibitem{}Bensby, T., Feltzing, S., Lundstr\"om, I., Ilyin, I.: 2004b, A\&A, 421, 969
\bibitem{}Bilir, S., Karaali, S., Gilmore, G.: 2005, MNRAS, (submitted)
\bibitem{}Carney, B.W., Latham, D.W., Laird, J.B.: 1990, AJ, 99, 572
\bibitem{}Chen, B., et al.: 2001, ApJ, 553, 184
\bibitem{}de Grijs, R., Peletier, R.F., van der Kruit, P.C.: 1997, A\&A, 327, 966
\bibitem{}de Vaucouleurs, G.: 1948, Ann. d'Astrophys., 11, 247
\bibitem{}Eggen, O.J., Lynden-Bell, D., Sandage, A.R.: 1962, ApJ, 136, 748
\bibitem{}Feltzing, S., Bensby, T., Lundstr\"om, I.: 2003, A\&A, 397, L1-L4
\bibitem{}Freeman, K., Bland-Hawthorn, J.: 2002, ARA\&A, 40, 487
\bibitem{}Gilmore, G., Reid, N.: 1983, MNRAS, 202, 1025
\bibitem{}Gilmore, G., Wyse, R.F.G.: 1985, AJ, 90, 2015
\bibitem{}Jahreiss, H., Wielen, R.: 1997, in: HIPPARCOS'97. Presentation
   of the HIPPARCOS and TYCHO catalogues and first astrophysical results of the
   Hipparcos space astrometry mission., Battrick, B. et al.  (eds.), ESA SP-402, 
   Noordwijk, p.675
\bibitem{}Karaali, S., Bilir, S., Hamzao\u glu\, E.: 2004, MNRAS, 355, 307
\bibitem{}Karaali, S., Bilir, S., Ak, S., Karata\c{s}, Y., Hamzao\u glu\, 
   E.: 2005 (in preparation)
\bibitem{}Kuijken, K., Gilmore, G.: 1989, MNRAS, 239, 571
\bibitem{}Nissen, P.E.: 2003, in McWilliam, A., Rauch, M., eds. Carnegie 
   Observatories Astrophysics Series, Vol.4
\bibitem{}Norris, J.E., Bessell, M.S., Pickles, A.J.: 1985, ApJS, 58, 463
\bibitem{}Norris, J.E., 1986: ApJS, 61, 667
\bibitem{}Norris, J.E., Ryan, S.G.: 1991, ApJ, 380, 403
\bibitem{}Norris, J.E., 1996: in: Formation of the Galactic halo inside and out, 
   ASP Conference Series, Vol. 92, 1996, Morrison, H. \& Sarajedini, A. (eds.), p.14
\bibitem{}O'Connell, D.J.K., ed. 1958. Stellar Populations. Amsterdam: North 
   Holland Press. Oort J.H., 1958. In O'Connell, p. 419  
\bibitem{}Phleps, S., Meisenheimer, K., Fuchs, B., Wolf, C., 2000, A\&A, 356, 108
\bibitem{}Robin, A., Cr\'{e}z\'{e}, M.: 1986, A\&A, 157, 71
\bibitem{}Sandage, A., Fouts, G.: 1987, AJ, 93, 74
\bibitem{}Searle, L., Zinn, R.: 1978, ApJ, 225, 357
\bibitem{}Siegel, M.H., Majewski, S.R., Reid, I.N., Thompson, I.B.: 2002, ApJ, 578, 151
\bibitem{}van der Kruit, P.C., Searle, L.: 1981a, A\&A, 95, 105
\bibitem{}van der Kruit, P.C., Searle, L.: 1981b, A\&A, 95, 116
\bibitem{}van der Kruit, P.C., Searle, L.: 1982a, A\&A, 110, 61
\bibitem{}van der Kruit, P.C., Searle, L.: 1982b, A\&A, 110, 79
\bibitem{}van der Kruit, P.C.: 1988, A\&A, 192, 117 
\bibitem{}Yoshii, Y., Saio, H.: 1979, PASJ, 31, 339
\bibitem{}Young, P.J.: 1976, AJ, 81, 807 
\end{thebibliography}
\end{document}